\documentclass[aps,prb,twocolumn,amsmath,amssymb,superscriptaddress,floatfix,citeautoscript,longbibliography]{revtex4-2}
\usepackage{graphicx} 
\usepackage{amsmath}
\usepackage{mathtools}
\usepackage{multirow}
\usepackage{xcolor}
\usepackage{bm}
\usepackage{amsfonts,amssymb,amsmath}
\usepackage{graphicx,dcolumn,bm,xcolor,braket,slashed}
\usepackage{times} 
\usepackage{comment}
\usepackage{array}
\usepackage{textcomp}
\usepackage[normalem]{ulem}
\usepackage{dsfont}
\usepackage{hyperref}

\begin{document}

\title{Klein tunneling in quantum geometric semimetals}
\author{Sang-Hoon Han}
\affiliation{Department of Physics, Ajou University, Suwon 16499, Republic of Korea}
\author{Jun-Won Rhim}
\email{jwrhim@ajou.ac.kr}
\affiliation{Department of Physics, Ajou University, Suwon 16499, Republic of Korea}
\author{Chang-geun Oh}
\email{cg.oh.0404@gmail.com}
\affiliation{Department of Applied Physics, The University of Tokyo, Tokyo 113-8656, Japan}
\date{Jun 2025}

\begin{abstract}
Klein tunneling stands as a fundamental probe of relativistic quantum transport in two-dimensional materials. 
We investigate this phenomenon in quadratic band-touching systems, where the Hilbert–Schmidt quantum distance plays a central role in the underlying mechanism.
By employing a generic parabolic model, we systematically disentangle the cooperative effects of intrinsic mass asymmetry and tunable quantum geometry. 
We demonstrate that mass asymmetry sets the overall transmission profile, including the angular distribution and the resonance channels.
In contrast, we show that quantum geometry provides a universal parameter that modulates tunneling efficiency by tuning the quantum distance, while leaving the energy dispersion unchanged.
Specifically, quantum geometry plays a dual role: it governs the overall transmission amplitude through pseudospin mismatch, while its interplay with Fabry–Pérot interference induces observable shifts in resonance angles. Our findings reveal that incorporating quantum geometry alongside band structure is essential for a complete description of quantum transport.
\end{abstract}

\maketitle

\section{Introduction}
Klein tunneling, the counterintuitive phenomenon of relativistic particles perfectly transmitting through high potential barriers, has long been a fascinating subject in quantum transport, first explored in the context of the Dirac equation~\cite{klein1929reflexion,su1993barrier,dombey1999seventy,calogeracos1999history}.
The advent of monolayer graphene, whose low-energy excitations behave as massless Dirac fermions, provided the first experimental platform to observe this effect~\cite{katsnelson2006chiral,stander2009evidence}.
In monolayer graphene, the pseudospin of each carrier is rigidly locked to its momentum, so that at normal incidence an electron outside the barrier and a hole inside share identical pseudospin states, yielding unity transmission ($T=1$) regardless of barrier height~\cite{katsnelson2006chiral,stander2009evidence,allain2011klein,castro2009electronic,ando1998berry}.
In contrast, bilayer graphene exhibits qualitatively different behavior. Its low-energy quasiparticles acquire an effective mass and a $2\pi$ Berry phase \cite{mccann2006electronic,novoselov2006unconventional,McCannKoshino2013}, so that an incident electron’s pseudospin becomes orthogonal to the corresponding hole state within the barrier at normal incidence, leading to complete suppression of transmission ($T=0$), which is known as anti-Klein tunneling~\cite{varlet2014fabry,kleptsyn2015chiral,katsnelson2006chiral}.
This anti–Klein tunneling is lifted at oblique incidence, where the transmission becomes finite and can be tuned, leading to a markedly different angular dependence of quantum transport\cite{tudorovskiy2012chiral,du2018tuning,dell2018klein,park2011pi,varlet2016band,liu2026mode}.

This sharp difference of the Klein tunneling between two semimetals highlights how profoundly the underlying electronic structure governs quantum transport \cite{mandal2020tunneling}.
Bilayer graphene is a prototypical example of a system hosting a quadratic band-touching (QBT) \cite{mccann2006electronic,McCannKoshino2013}. 
However, it represents a rather specific case within the family of QBT semimetals, in which the conduction and valence bands are characterized by symmetric effective masses.
The family of QBT materials is, in fact, far more diverse and includes systems with intrinsically asymmetric effective masses, such as in monolayers of phosphorene oxide~\cite{zhu2016blue} and magnesium carbide~\cite{wang2018monolayer}, or in certain kagome lattices~\cite{park2020kagome}. 
They share the same pseudospin structure with bilayer graphene but possess asymmetric effective masses~\cite{oh2025universal}.
Furthermore, recent studies have revealed that the physics of QBT systems is dictated not only by the band dispersion but also by the underlying quantum geometry of the electronic wavefunctions, such as 
bulk-interface correspondence~\cite{oh2022bulk,kim2023general}, optical properties~\cite{oh2025universal,oh2025color}, unconventional Landau level structures~\cite{rhim2020quantum,oh2024revisiting}, magnetic properties~\cite{oh2025magnetic}, and enhanced thermoelectric efficiency \cite{oh2024thermoelectric}.
Studies dealing with quantum geometry like these are currently among the main research topics actively pursued in condensed matter physics \cite{yu2025quantum,liu2025quantum,asteria2019measuring,yu2024non,komissarov2024quantum,gianfrate2020measurement,kang2025measurements,tanaka2025superfluid,oh2025role}.

Despite the significant progress in understanding the geometric effects in QBT systems, their direct consequences on fundamental transport phenomena like Klein tunneling are not yet fully understood.
Specifically, how the interplay between tunable wavefunction geometry and the intrinsic mass asymmetry—a feature absent in bilayer graphene—cooperatively reshapes the tunneling process remains an open question. 

In this paper, we bridge this gap by theoretically investigating Klein tunneling in a generalized isotropic QBT model~\cite{oh2025universal}. 
We systematically analyze how the transmission probabilities are governed by the combined effects of wavefunction geometry and mass asymmetry, providing a comprehensive framework for quantum transport in this diverse class of materials having QBTs.
Our investigation shows that mass asymmetry governs the angular dependence of transmission, shaping both the number and the angular positions of the resonance peaks.

In the presence of mass asymmetry, we isolate the intrinsic geometric degree of freedom encoded in the wavefunction. We find that the quantum geometry governs the pseudospin mismatch between the incident electron and the hole state in the barrier region, thereby determining the transmission amplitude.
Specifically, a reduction in this wavefunction-geometric nontriviality leads to a suppression of the overall tunneling efficiency.

To rigorously validate the intrinsic origin of this suppression, we derive the exact analytical transmission amplitude for a delta-function barrier. This solution explicitly reveals the dependence of the transmission amplitude on the quantum geometry. Ultimately, our findings provide a clear roadmap for band geometry engineering, establishing quantum geometry as a practical tool for manipulating carrier transport.

\section{General Isotropic QBT model}

\begin{figure}[t]
    \centering
    \includegraphics[width=86mm]{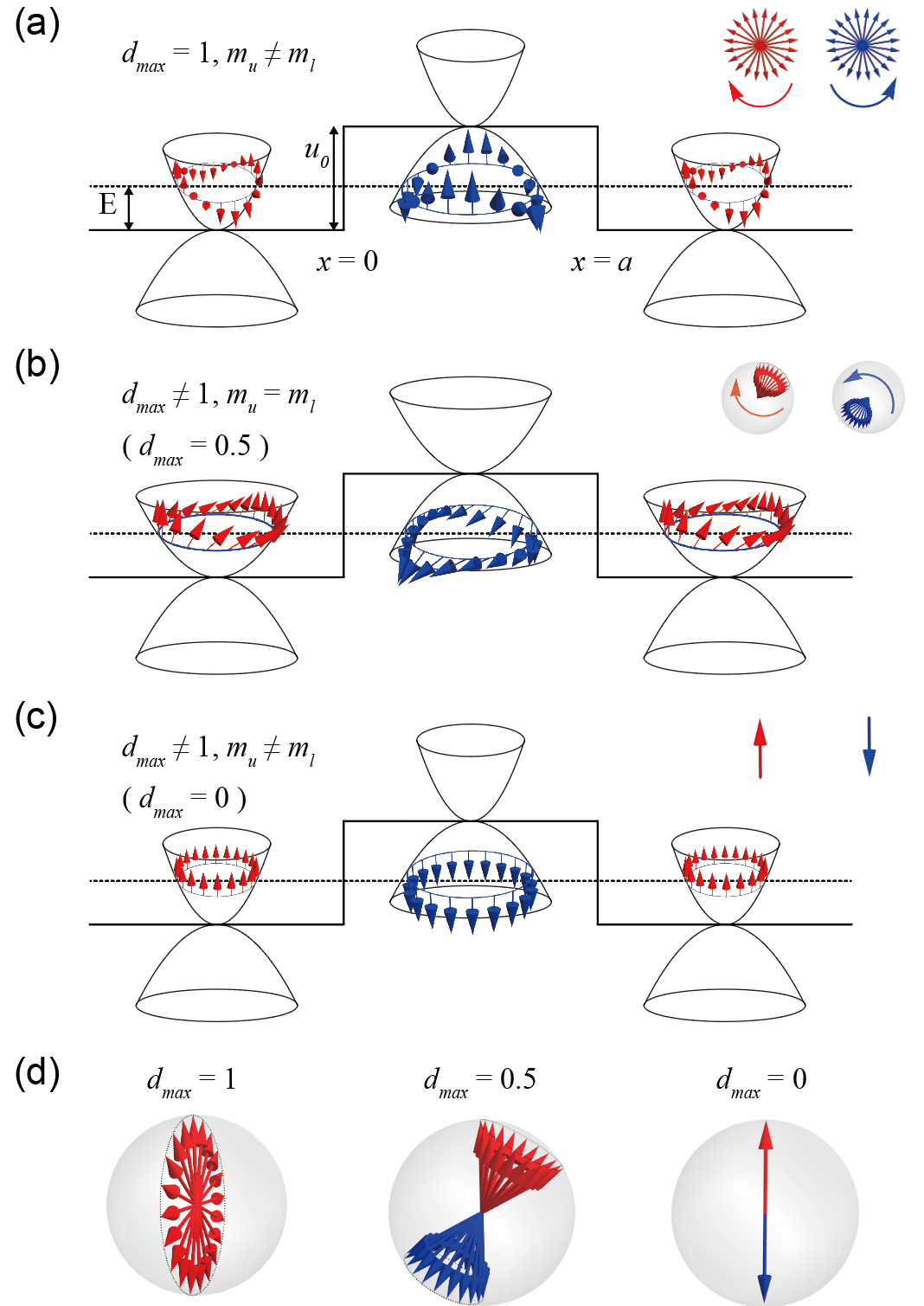}
    \caption{
Schematic illustration of quantum tunneling and pseudospin textures in a QBT system. 
Pseudospin orientation of propagating quasiparticle states is plotted for three distinct regions: incident ($x<0$), barrier ($0<x<a$), and transmitted ($a<x$). 
The red arrows indicate the direction of the pseudospin. The barrier has height $u_0$ and width $a$; $E$ is the incoming particle energy. 
The pseudospin of the upper band rotates in the opposite direction compared to that of the lower band.
(a) Mass asymmetric system with $d_{\mathrm{max}}=1$. 
(b) Mass symmetric system with $d_{\mathrm{max}}=0.5$.
(c) Mass asymmetric system with $d_{\mathrm{max}}=0$. The insets in (a-c) plot the pseudospin orientation as a function of the incidence angle for each corresponding $d_{\mathrm{max}}$ value.
(d) Enlarged plot of the pseudospin orientation as a function of incidence angle for $d_{\mathrm{max}}=1, 0.5, \text{and } 0$. Red and blue arrows indicate the pseudospin for the upper and lower bands, respectively. For $d_{\mathrm{max}}=1$, the pseudospins of both bands are aligned (overlap), so only the red arrow is shown.
}
 \label{fig:fig1}
\end{figure}
We consider a two-dimensional continuum model for a system with a QBT point between two isotropic bands. The energy dispersion for the upper ($u$) and lower ($l$) bands is given by
\begin{align} \epsilon_{u/l}(\bm{k}) =\frac{ \hbar^2k^2}{2 m_{u/l}}, \label{eq:dispersion} \end{align}
where $\bm{k}=(k_x, k_y)$ is the momentum, $k^2=\bm{k}\cdot\bm{k}$, and the effective masses $m_u$ and $m_l$ can be positive, negative, or infinite, corresponding to different band structures.
The most general two-band Hamiltonian for such a system is parameterized by three parameters: two effective masses $m_u$, $m_l$, and an interband coupling $d$ \cite{oh2024revisiting,oh2025universal}. 
It can be written as 
    \begin{eqnarray}
        \mathcal{H}_0(\bm{k}) = \hbar^2\sum_{\gamma } h_\gamma (\bm{k}) \sigma_\gamma , \label{eq:Ham}
    \end{eqnarray}
where $\sigma_\gamma$ represents the identity ($\gamma=0$) and Pauli matrices ($\gamma = x,y,z$). 
Additionally, $h_\gamma (\bm{k})$ is a real quadratic function: $h_{0} (\bm{k}) = (1/M+2/m_l)(k_x^2+ k_y^2)/4$, $h_{x} (\bm{k}) =d \sqrt{1-d^2} k_y^2/(2M),~h_{y} (\bm{k}) =  d k_x k_y/(2M)$, and $h_{z} (\bm{k}) = \left[k_x^2+(1-2d^2)k_y^2\right]/(4M)$, where $1/M = 1/m_u -1/m_l$. 
In the definition of $h_x(\bm{k})$, the parameter $d$ is implicitly assumed to range from $-1$ to $1$.

The parameter \(d\) itself encapsulates the quantum geometry.
The Hilbert–Schmidt quantum distance between eigenstates across momentum space is defined by
\begin{equation}
d^2_{\mathrm{HS},nm}(\bm{k}, \bm{k}') = 1 - \left| \langle u_{n,\bm{k}} | u_{m,\bm{k}'} \rangle \right|^2,
\end{equation}
where \( u_{n,\bm{k}} \) denotes the eigenstate of band \( n \) at momentum \( \bm{k} \).
For the QBT semimetals, one of the key quantities governing the low-energy physics is the maximum value of $d_{\mathrm{HS},nn}(\bm{k}, \bm{k}')$ between eigenvectors around the touching point, denoted by $d_\mathrm{max}$.
Note that $d_\mathrm{max}$ is the same for the upper and lower bands.
Both $d_{\mathrm{HS},nm}(\bm{k}, \bm{k}')$ and $d_\mathrm{max}$ values are from $0$ to $1$.
We introduce an associate quantum distance $d=\xi d_\mathrm{max}$, where $\xi = \pm1$.
Here, \(\xi=\pm1\) is determined by a valley index or other discrete degrees of freedom.
Without loss of generality, we set $\xi=+1$ hereafter, as its sign does not affect the Klein tunneling characteristics.
Note that we consider the isotropic energy dispersion (i.e., $\epsilon_{u/l}\propto k^2$), which does not require the Hamiltonian itself to be rotationally symmetric.
While the geometric quantity $d_\mathrm{max}$ represents the strength of the inter-band coupling, it does not manifest in the band dispersion, as the band dispersion is entirely determined by only two parameters ($m_u, m_l$).

We consider electron tunneling through a square potential barrier of height ($u_0$) and width ($a$), applied along the x-direction [Fig.~\ref{fig:fig1}]:
\begin{equation}
V(x) =
\begin{cases}
u_0, & 0 < x < a, \\
0, & \text{otherwise}.
\end{cases}
\end{equation}
Due to the translational invariance along the y-direction, the transverse momentum ($k_y$) is conserved. 
In each of the three regions ($i \in \{1, 2, 3\}$) (region 1: ($x<0$), region 2: ($0<x<a$), region 3: ($x>a$)), we therefore write the spinor wavefunction as a superposition of propagating and evanescent modes,
\begin{align}
\Psi^{(i)}(x, y) = (&a_i A_i e^{i k_{ix} x} + b_i B_i e^{-i k_{ix} x} \nonumber \\
&+ c_i C_i e^{\kappa_{ix} x} + d_i D_i e^{-\kappa_{ix} x}) e^{i k_y y}
\end{align}
where $\hbar\bm{k}_i = \sqrt{2m_i|E-V_i|}(\cos\phi_i,\sin\phi_i)$, $\kappa_{ix} = \sqrt{k_{ix}^2 + 2k_{iy}^2}$, $E$ is kinetic energy of the incident electron and $V_i$ represents the potential in the $i$-th region.

The basis spinors ($A_i, B_i, C_i, D_i$) define the intrinsic structure of the eigenstates in region $i$.
The scalar coefficients ($a_i, b_i, c_i, d_i$) represent the overall scattering amplitudes for the propagating and evanescent modes associated with these basis spinors in each region. We determine these amplitudes by applying continuity boundary conditions for the total spinor wavefunction $\Psi^{(i)}(x,y)$ at the interfaces. From these, the transmission probability is calculated as $T = |a_3/a_1|^2$ for transmission from region 1 to 3. The effective masses in different regions are denoted by $m_i$. For $E>V_i$, $m_i = m_u$, while for $E<V_i$, $m_i=m_l$.
Full expressions for the wavefunction components and matching conditions are provided in Appendix~\ref{app:bc}.

\begin{figure}[t]
    \centering
    \includegraphics[width=86mm]{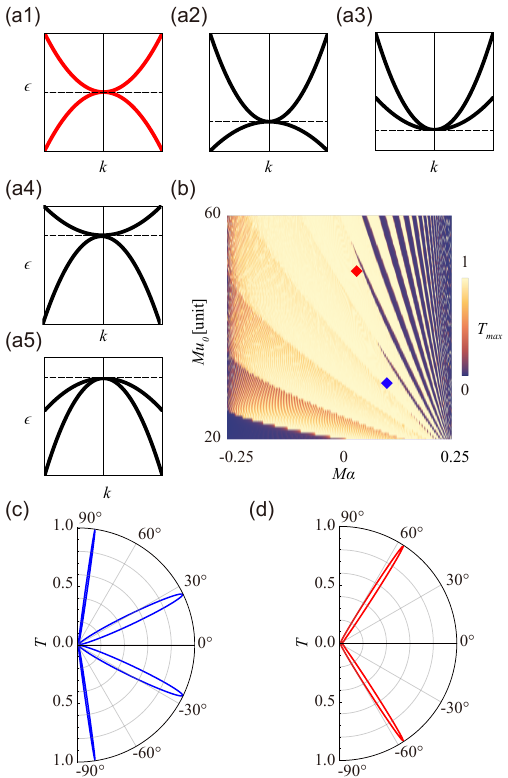}
    \caption{
    (a) QBT band structures for different values of $M\alpha$, where $M\alpha$ is the mass asymmetry, which is product of the mass asymmetry parameter $\alpha$ and the relative effective mass $M$:
    (a1) \( M \alpha = 0 \),  
    (a2) \( 0 < M \alpha < 1/4 \),  
    (a3) \( 1/4 \leq M \alpha \),
    (a4) \( -1/4 < M \alpha < 0 \), 
    (a5) \( M \alpha \leq -1/4 \). The dashed line represents the zero-energy line ($\varepsilon=0$).
    (b) Contour plot of the transmission probability $T_{\mathrm{max}}$ (representing the maximum value of $T$ across all incidence angles) as a function of mass asymmetry parameter $M \alpha$ and the barrier height $M u_0$. Blue and red dots indicate examples of unit transmission ($T_{\mathrm{max}}=1$) points shown in Fig. 2(c) and (d). Here, [unit] = $[m_\mathrm{BLG}/2 \cdot \mathrm{meV}]$ with the effective mass of bilayer graphene $m_{\mathrm{BLG}} =2.0122 \times 10^{-10}(\text{meVs}^2/\text{m}^2)$.
    (c, d) Transmission probability \( T \) as a function of incidence angle \( \phi \) for (c) $(M \alpha,Mu_0)=  (0.102,30~[\text{unit}])$ and (d) $ (M \alpha,Mu_0)=  (0.035,50~[\text{unit}])$, corresponding to the blue and red dots in Fig. 2(b), respectively.
    These results are calculated for $ME = 17~[\text{unit}]$ and $a = 100~ \text{nm}$.
       }   
    \label{fig:fig2}
\end{figure}

\section{asymmetric effective masses}
The general QBT model presented in the previous section contains a particularly important special case: the mass-symmetric case $m_u=-m_l$.
When combined with $d_\mathrm{max}=1$, this limit is unitarily equivalent to the low-energy bilayer-graphene Hamiltonian.
Specifically, in pristine bilayer graphene, this symmetric mass condition with the nontrivial wavefunction structure leads to unique transport phenomena, most notably the complete suppression of transmission at normal incidence (anti-Klein tunneling) and the emergence of perfect transmission resonances at finite oblique angles \cite{katsnelson2006chiral,tudorovskiy2012chiral}.
These phenomena are intrinsic consequences of the specific band structure and pseudospin winding associated with the $d_{\mathrm{max}}=1$ geometry. As noted in the introduction, this system serves as the standard theoretical and experimental benchmark for Klein tunneling. Crucially, for bilayer graphene, the $C_3$ rotational and time-reversal symmetries pin the quantum distance parameter to $d_{\mathrm{max}}=1$~\cite{oh2025universal,hwang2021wave}, rendering it a distinct, symmetry-protected realization of the QBT model.

Although bilayer graphene is the prototypical isotropic QBT with $d_{\mathrm{max}}=1$, its perfect mass symmetry is not universal.
A diverse class of materials, including monolayers of PO~\cite{zhu2016blue}, Na$2$O~\cite{feng2021two}, and Mg$2$C~\cite{wang2018monolayer}, are predicted to host isotropic QBT points protected by the same $C_3$ and time-reversal symmetries, thereby sharing the $d_{\mathrm{max}}=1$ constraint~\cite{oh2025universal}.
However, unlike bilayer graphene, these materials are expected to exhibit asymmetric effective masses ($|m_u| \neq |m_l|$).
Figure~\ref{fig:fig2}(a) categorizes these mass-asymmetric scenarios.
This material landscape offers a unique opportunity to isolate the impact of mass asymmetry on quantum transport, as the geometric parameter $d{\mathrm{max}}$ remains fixed across this class.

To systematically explore mass asymmetry, we analyze the Hamiltonian in Eq.~(\ref{eq:Ham}) with fixed $d_{\mathrm{max}}=1$.
We quantify the asymmetry using the parameter $\alpha \equiv h_0/k^2 = (1/M+2/m_l)/4 = (1/m_u+1/m_l)/4$.
This parameter is strictly zero for bilayer graphene ($\alpha=0$), with non-zero values indicating deviation from mass symmetry.

The parameter $\alpha$ delineates distinct dispersion regimes of the band structure, as illustrated in Figs.~\ref{fig:fig2}(a1--a5):
\begin{itemize}
\item Symmetric Case ($\alpha = 0$): The system reduces to the standard mass-symmetric QBT model (a1).
\item {Asymmetric Klein Tunneling Regime ($0 < |M\alpha| < 1/4$):} The bands exhibit unequal curvatures but opposite signs. Note that the band structure for $M\alpha < 0$ (a4) is the energy-inverted counterpart of the $M\alpha > 0$ case (a2).
\item {Blocked Regime ($|M\alpha| \geq 1/4$):} A transition occurs where both effective masses acquire the same sign (a3, a5). In this regime, the electron–hole band structure for Klein tunneling is lost, blocking the transmission pathway.
\end{itemize}

Klein tunneling is governed by multiple factors, including mass asymmetry ($M\alpha$), quantum geometry ($d_\mathrm{max}$), barrier height ($Mu_0$), barrier width ($a$), and incident energy ($ME$). 
In this section, we characterize the transport response by the maximum transmission $T_{\max}$ and construct a two-parameter transmission map in the $(M\alpha, Mu_0)$ plane, while keeping $ME$ and $a$ fixed (the sensitivity to other parameters is discussed in Appendix~\ref{app:parameters}). 
Here, $T_{\max}$ denotes the maximum of the angle-resolved transmission over the incident angle at the specified ($ME, a, M\alpha, Mu_0$).
Figure~\ref{fig:fig2}(b) presents a color map of the maximum transmission probability, $T_{\mathrm{max}}$, as a function of mass asymmetry $M\alpha$ and barrier height $Mu_0$.
Each point on the map represents the maximum transmission probability found across all possible incidence angles for a given pair of ($M\alpha, Mu_0$).
The incident energy and barrier width are chosen as $ME = 17~[\text{unit}]$, where $[\text{unit}] = m_{\mathrm{BLG}}/2 \cdot \mathrm{meV}$, and $a = 100~ \text{nm}$ , aligning with the standard bilayer graphene literature~\cite{katsnelson2006chiral} to facilitate direct comparison.

A striking feature of the map is the sharp vanishing of transmission ($T_{\mathrm{max}} \approx 0$) even within the propagation-allowed window ($|M\alpha| < 0.25$).
This behavior is best elucidated by the electron analogy of Snell's Law, 
\begin{align}
k_1 \sin \phi = k_2 \sin \theta.
\end{align}
Using this optical framework, we explain two distinct mechanisms driving the suppression near the transition line ($|M\alpha| = 1/4$).
The first mechanism is the refractive narrowing of the transmission window. As $M\alpha$ approaches $-1/4$, the condition $k_1 \gg k_2$ leads the critical angle for total internal reflection, $\phi_c = \arcsin(k_2/k_1)$, to shrink toward zero. This restricts electron entry to a narrow cone near normal incidence. Crucially, since transmission at $\phi=0$ is forbidden by pseudospin orthogonality (anti-Klein tunneling), the available transport channels are effectively pinched off.

The second mechanism involves the suppression of Fabry-Pérot (FP) resonances. These resonances arise from the coherent interference of quasiparticle waves undergoing multiple reflections between the two barrier interfaces \cite{shytov2008klein,varlet2014fabry,young2009quantum}. Perfect transmission ($T=1$) occurs when this interference is constructive, which requires satisfying a specific phase condition at a target refraction angle $\theta_{\mathrm{res}}$. The corresponding incidence angle needed to access this internal resonance is given by $\sin \phi_{\mathrm{res}} = (k_2/k_1) \sin \theta_{\mathrm{res}}$. As $M\alpha$ approaches $1/4$ ($k_2 \gg k_1$), the required $\sin \phi_{\mathrm{res}}$ can exceed unity for all available $\theta_{\mathrm{res}}$. In this regime, Snell’s law admits no physical incidence angle for $\phi_{\mathrm{res}}$, meaning the would-be resonant channel no longer corresponds to a physical scattering state. The FP resonance thus becomes physically inaccessible.

From an FP perspective, the role of mass asymmetry can be rationalized in terms of the internal dispersion. 
In the simplest FP limit, perfect transmission occurs at incidence angles satisfying
\begin{align}
k_{2x} a = \pi n,
\end{align}
where $n$ is an integer. 
In monolayer graphene, where the Dirac spinor structure is particularly simple, this condition indeed coincides with the exact resonance angles of the standard barrier problem~\cite{allain2011klein,katsnelson2006chiral}. 
In our case, however, quantum geometric corrections encoded in the interface reflection phases shift the exact angles at which $T=1$ is reached, even though the FP condition still tracks the resonance angles to good accuracy.

Satisfying the FP condition does not, by itself, guarantee unit transmission, since residual quantum geometric effects can still reduce the transmission amplitude away from unity. 
Nevertheless, the FP condition remains a useful predictive framework for identifying resonance points: for fixed energy and barrier parameters, it defines a discrete set of candidate resonance angles $\{\phi_n(a)\}$ through $2 q_x(\phi_n) a = 2\pi n$. 
As the barrier width $a$ increases, the number of such FP resonance angles within the propagation-allowed window grows, providing a natural interpretation of the numerical trend that the parameter region with $T_{\mathrm{max}} = 1$ expands at larger $a$, as seen in Appendix~C.

Beyond the maximum transmission, the full angular profile $T(\phi)$ reveals richer structures. As shown in Figs.~\ref{fig:fig2}(c,d), the number and position of transmission resonances are sensitive to the specific combination of ($M\alpha, Mu_0$).
We note that all profiles satisfy the symmetry $T(\phi) = T(-\phi)$, a necessary consequence of the even parity of the energy dispersion $\epsilon(\bm{k})$ with respect to $k_y$ and the transverse symmetry of the potential barrier.


\begin{figure}[t]
    \centering
    \includegraphics[width=86mm]{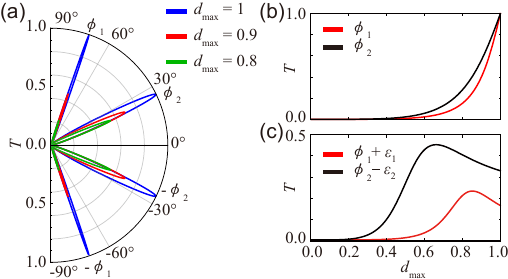}
    \caption{
    (a) Transmission probability \( T \) as a function of incidence angle \( \phi \) for \( d_\mathrm{max} = 1 \) (blue), \( 0.9 \) (red), and \( 0.8 \) (green), respectively. The angles \( \phi_1 \) and \( \phi_2 \) indicate the incidence directions at which perfect transmission occurs for \( d_\mathrm{max} = 1 \).
    (b) Transmission probability \( T \) at two representative incidence angles, \( \phi_1 \) (red) and \( \phi_2 \) (black). As \( d_\mathrm{max} \) decreases, the transmission is suppressed in both cases, but with distinct decay profiles, indicating angle-dependent sensitivity to the underlying quantum geometry.
    (c) Transmission probability $T$ as a function of $d_{\mathrm{max}}$ at two incidence angles near the representative perfect transmission angles: $\phi_1 + \epsilon_1$ (red) and $\phi_2 - \epsilon_2$ (black). Here, $\epsilon_1 = 0.03 \phi_1$ and $\epsilon_2 = 0.2 \phi_2$. 
    These results are calculated for $M\alpha=0$, $Mu_0=50~[\text{unit}]$, $ME = 17~[\text{unit}]$ and $a = 100~ \text{nm}$. }
 \label{fig:fig3}
\end{figure}

\section{Role of $d_\mathrm{max}$ in the symmetric mass limit}
Complementing the analysis of mass asymmetry, we now isolate the intrinsic influence of quantum geometry by tuning the parameter $d_\mathrm{max}$ within the symmetric mass limit ($\alpha=0$).
Figure~\ref{fig:fig3}(a) illustrates the angular dependence of the transmission probability, $T(\phi)$, for varying geometric parameters: $d_{\mathrm{max}} = 1$ (blue), $0.9$ (red), and $0.8$ (green).
A reduction in $d_\mathrm{max}$ leads to a general suppression of the transmission profile, although the angular structure remains nontrivial.
To quantify this suppression at the primary transmission channels, Fig.~\ref{fig:fig3}(b) tracks $T$ as a function of $d_{\mathrm{max}}$ at two representative angles, $\phi_1$ and $\phi_2$, which correspond to perfect transmission resonances for the $d_\mathrm{max}=1$ case.
At these key angles, we observe that the transmission decreases monotonically as $d_\mathrm{max}$ is tuned from unity to zero, suggesting a close connection between the tunneling efficiency and the wavefunction geometry.

This behavior finds a natural interpretation in the pseudospin formalism, as visualized in Fig.~\ref{fig:fig1}.
In a two-band model, the geometric properties are encoded in the texture of the pseudospin field $\bm{s}_{\bm k}=\bra{+\bm k} \bm{\sigma} \ket{+\bm k}$.
The quantum distance serves as a measure of the misalignment between these pseudospins: $d_{\mathrm{HS},nn}^2 (\bm{k} , \bm{k'})=(1-\bm{s}_{\bm{k}} \cdot \bm{s}_{\bm{k'}})/2$. Consequently, the parameter $d_\mathrm{max}$ dictates the overall ``canting" strength of the pseudospin texture.
The $d_{\mathrm{max}}=1$ limit (pristine bilayer graphene) is characterized by a winding number of 2~\cite{McCannKoshino2013}, meaning the pseudospin orientation rotates twice as fast as the momentum angle [Figs.~\ref{fig:fig1}(a) and (d)].
Crucially, the pseudospins of the electron and hole bands counter-rotate relative to each other. 
This distinct pseudospin behavior ensures the existence of specific oblique angles where the incident electron's pseudospin perfectly aligns with that of the hole state inside the barrier, despite their belonging to different bands.
This perfect alignment completely suppresses backscattering, resulting in perfect transmission ($T=1$).
While the underlying mechanism—pseudospin conservation—is analogous to the perfect transmission in single-layer graphene, the higher winding number in this system shifts the resonance condition from normal incidence to finite angles \cite{katsnelson2006chiral,allain2011klein,varlet2016band}.

As $d_{\mathrm{max}}$ is reduced, the pseudospin field undergoes a gradual reorientation [Fig.~\ref{fig:fig1}(b,d)]
This reorientation increases the intrinsic mismatch between the pseudospin states across the potential barrier interfaces, thereby diminishing the wavefunction overlap and suppressing transmission.
In the trivial limit of $d_\mathrm{max}=0$, all pseudospins collapse onto a single direction [Fig.~\ref{fig:fig1}(c)], rendering the overlap minimal and yielding zero transmission at all angles.

We corroborate this intuitive picture by analytically computing the wavefunction overlap between the incident electron state (u) and the transmitted hole state (l). The squared overlap amplitude is derived as
\begin{equation}
|\langle u_{u,\bm{k}} | u_{l,\bm{q}} \rangle|^2 = 1-d^2_{\mathrm{HS},ul}(\bm{k},\bm{q})=\frac{d_\mathrm{max}^2 \, k_y^2 (k_x - q_x)^2}{(k_x^2 + k_y^2)(q_x^2 + k_y^2)},
\end{equation} 
where $\bm{k}$ and $\bm{q}$ denote the momentum of the incident electron and transmitted hole states, respectively.

Expressing this in terms of the incidence $\phi$ and refraction $\theta$ angles yields 
\begin{equation}
|\langle u_{u,\bm{k}} | u_{l,\bm{q}} \rangle|^2  = d_\mathrm{max}^2 \sin^2 \phi \, \frac{(k \cos \phi - q \cos \theta)^2}{q^2}.
\end{equation}
These expressions explicitly show that the overlap is proportional to $d_\mathrm{max}^2$. 
This scaling is consistent with the interpretation that the tunneling suppression at the resonance angles $\phi_{1,2}$ arises from an enhanced pseudospin mismatch induced by the quantum geometric modification.
Collectively, these findings demonstrate that quantum geometry acts as a fundamental modulator of transport, suggesting that tunneling probabilities can be actively tuned via band geometry engineering.

While the reduction of $d_\mathrm{max}$ generally suppresses the overall transmission envelope, the detailed angular profile reveals a rich phenomenology.
As highlighted in Fig.~\ref{fig:fig3}(c), the transmission at specific angles can exhibit non-monotonic behavior, locally increasing as $d_\mathrm{max}$ decreases.
This counter-intuitive feature signifies a competitive interplay between two distinct mechanisms: the pseudospin mismatch (controlled by $d_\mathrm{max}$), which suppresses transmission, and 
FP resonance, which enhances transmission via constructive interference.
This interference generates a complex oscillatory pattern dependent on the barrier width $a$ and the in-barrier wavevector~\cite{varlet2014fabry,varlet2016band}.
Since the geometric parameter modifies the effective boundary conditions, tuning $d_\mathrm{max}$ shifts the resonance angles. Consequently, at angles where the resonance condition becomes satisfied due to this shift, the constructive interference can locally override the geometric suppression, leading to the complex peak shifts and crossings observed in Figs.~\ref{fig:fig3}(a) and (c).

\section{Intrinsic Geometric Effects in delta-function barrier}
In the finite-width barrier ($a>0$) regime considered thus far, the transmission spectrum is inevitably modulated by FP resonances, which can obscure the pure, intrinsic geometric effects governed by the single-interface scattering properties.
To disentangle the intrinsic role of quantum geometry from these extrinsic resonance complexities, we examine the delta-function barrier limit ($a\to0$ while keeping the barrier strength $U=u_0 a$ fixed). In this limit, the phase accumulation within the barrier vanishes, leaving the transmission to be determined solely by the wavefunction mismatch at the interface. The transmission amplitude simplifies to the exact compact form:
\begin{equation}
t(\phi; d_\mathrm{max})=\frac{1}{1+iA(\phi)+d_\mathrm{max}^{2} B(\phi),},
\label{eq:delta-t-compact}
\end{equation}
where $A(\phi)$ represents the FP-free baseline impedance (determined by the barrier strength and incidence angle) and $B(\phi)$ serves as the geometry kernel. 
This analytical result offers two crucial insights. 
First, it rigorously validates the intuition derived from spinor overlaps: the geometric contribution is cleanly isolated and enters the transmission amplitude exclusively through the $d_\mathrm{max}^{2} B(\phi)$ term.
Second, it allows for the derivation of an exact, closed-form condition for perfect transmission ($T=1$). By introducing the dimensionless barrier strength $\rho\equiv \sqrt{M} U/(\sqrt{E}\hbar)$, the angles of perfect transparency are determined by the roots of:
\begin{equation}
x\Big[x(x-3)+2\sqrt{2}\rho\sqrt{3-x}\Big]=2\rho^{2},
\label{eq:T1-delta}
\end{equation}
where $x \equiv \cos(2\phi)$. 
The detailed derivation of Eqs.~\eqref{eq:delta-t-compact} and \eqref{eq:T1-delta} is provided in Appendix~\ref{app:delta}. These resonance-free results establish an essential benchmark for interpreting the finite-width barrier results, where FP resonances further modulate the peak positions and widths.

\begin{figure}[t]
    \centering
    \includegraphics[width=86mm]{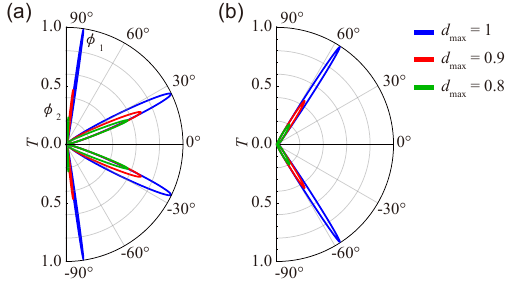}
    \caption{
    Transmission probability $T$ as a function of incidence angle $\phi$ for  $d_{\mathrm{max}} = 1$ (blue), $0.9$ (red), and $0.8$ (green), respectively. (a) $(M \alpha,Mu_0)=  (0.102,30~[\text{unit}])$ and (b) $ (M \alpha,Mu_0)=  (0.035,50~[\text{unit}])$, corresponding to Figs. 2(c) and (d), respectively.
    These results are calculated for $ME = 17~[\text{unit}]$ and $a = 100~ \text{nm}$. Here, $\phi_1$ and $\phi_2$ correspond to the transmission peak angles for $d_{\max}=1$ and $d_{\max}=0.8$, respectively.
    }
 \label{fig:fig4}
\end{figure}

\section{Interplay of Mass Asymmetry and Quantum Geometry}
Having separately analyzed the effects of mass asymmetry (governed by $M\alpha$) and quantum geometry (governed by $d_\mathrm{max}$), we now investigate their combined influence on Klein tunneling.
The central question is whether the geometric suppression mechanism identified in the symmetric limit persists in mass-asymmetric systems, and how these two control parameters jointly reshape the transmission landscape.

To explore this interplay, we revisit the representative mass-asymmetric configurations defined in Figs.~\ref{fig:fig2}(c) and (d) and track the evolution of their transmission profiles as the geometric parameter is detuned from the bilayer graphene limit ($d_\mathrm{max}=1$).
Figures~\ref{fig:fig4}(a) and (b) present the angular dependence of the transmission probability for the baseline case $d_\mathrm{max}=1$ (blue) alongside reduced values of $d_\mathrm{max}=0.9$ (red) and $d_\mathrm{max}=0.8$ (green).

The data reveal a clear hierarchy of control between the two parameters.
The mass asymmetry parameter $M\alpha$ constructs the fundamental propagation landscape of the transport; it dictates the global angular structure, establishing the allowable transmission windows and the approximate angular positions of the resonances.
Given this angular structure,
the quantum geometry ($d_\mathrm{max}$) acts as the primary modulator of transmission efficiency.
Consistent with the symmetric case, a reduction in $d_\mathrm{max}$ generally suppresses the transmission amplitude, suggesting that pseudospin mismatch remains the dominant suppression mechanism even in the presence of mass asymmetry.

Crucially, however, this geometric modulation is not merely a multiplicative scaling factor.
Because the geometric parameter $d_\mathrm{max}$ enters the boundary matching conditions, it effectively modifies the interfacial impedance seen by the electron. This modification renormalizes the FP resonance conditions, leading to observable shifts in the peak positions.
For instance, in Fig.~\ref{fig:fig4}(a), the transmission peak for $d_\mathrm{max} =0.8$ ($\phi_2$) is noticeably shifted towards $\pi/2$ relative to the $d_\mathrm{max}=1$ ($\phi_1$) case.
This demonstrates that while mass asymmetry sets the stage for transport, quantum geometry actively tunes both the amplitude and the precise angular phase of the tunneling resonances.


\section{Conclusion}
In conclusion, we have established a comprehensive framework for Klein tunneling in generalized QBT systems, demonstrating that transport characteristics are governed by a cooperative interplay between mass asymmetry and quantum geometry. Our findings reveal a distinct functional hierarchy between these two control parameters: mass asymmetry constructs the fundamental propagation landscape—defining the accessible angular ranges and the structure of resonance channels—while quantum geometry acts as the universal control knob that determines tunneling efficiency via the mechanism of pseudospin mismatch.
This geometric control is not merely qualitative but intrinsic, as we proved by deriving the explicit dependence of the transmission amplitude in the resonance-free delta-barrier limit.

Our work provides a clear roadmap for ``band geometry engineering", demonstrating that the quantum geometry of a material can be used as a practical tool to control its electrical conductance. 
For example, by invoking the Landauer formalism ($G\propto T$), which identifies conductance as the cumulative transmission probability of scattering channels~\cite{landauer1970electrical,fisher1981relation}, a direct physical link between the abstract geometric variations of the wavefunction and macroscopic transport observables can be established.
More broadly, geometry–driven control of the angular transmission profile offers a microscopic route to angle–selective transport, which underlies a variety of ballistic electron–optics phenomena—such as Veselago lensing \cite{park2011pi,hassler2010flat}, caustic focusing \cite{cserti2007caustics,peterfalvi2010catastrophe}, angle filtering in superlattice potentials\cite{barbier2008dirac,barbier2009bilayer}, and supercollimation \cite{park2008electron,barbier2010single}—and their photonic analogues~\cite{fang2016klein}.

\begin{acknowledgments}
 J.W.R was supported by the National Research Foundation of Korea (NRF) Grant funded by the Korean government (MSIT) (Grant nos. 2021R1A2C1010572 and 2022M3H3A1063074). C.O. was supported by JSPS KAKENHI Grant Number JP25KF0186. This research was supported by Global - Learning \& Academic research institution for Master’s·PhD students, and Postdocs(G-LAMP) Program of the National Research Foundation of Korea(NRF) grant funded by the Ministry of Education(No. RS-2023-00285390).
\end{acknowledgments}

\bibliographystyle{apsrev4-1}

\bibliography{references}
\onecolumngrid

\clearpage

\appendix
\setcounter{figure}{0}
\renewcommand{\thefigure}{\thesection\arabic{figure}}
\newpage
\section{Boundary Conditions and Transmission Probability}
\label{app:bc}

This Appendix provides the detailed theoretical framework for the derivation of the transmission probability discussed in the main text. We begin by introducing the QBT Hamiltonian that governs the system and defining the scattering problem setup, including the potential barrier and the resulting wave vectors. Subsequently, we establish the appropriate boundary conditions for the wavefunctions at each interface and outline the systematic procedure for matching them, which allows us to determine the coefficients of the reflected and transmitted waves. Finally, we show how the transmission probability is obtained from these coefficients by defining the probability current, providing the full theoretical framework supporting our main results.

In this paper, We use the isotropic QBT Hamiltonian~\cite{oh2025universal}
\(
H_0(\mathbf{k})=\sum_{\alpha=0,x,y,z} h_\alpha(\mathbf{k})\,\sigma_\alpha
\)
with
\[
h_0=\tfrac14\!\left(\tfrac{1}{M}+\tfrac{2}{m_l}\right)(k_x^2+k_y^2),\quad
h_x=\tfrac{d\sqrt{1-d^2}}{2M}\,k_y^2,\quad
h_y=\tfrac{d}{2M}\,k_xk_y,\quad
h_z=\tfrac{1}{4M}\!\left[k_x^2+(1-2d^2)k_y^2\right],
\]
where \(d\) is the geometric parameter and
\[
\frac{1}{M}=\frac{1}{m_u}-\frac{1}{m_l}.
\]
The band dispersions are \(\varepsilon_{u/l}(\mathbf{k})=\mathbf{k}^2/(2m_{u/l})\) and thus depend only on \((m_u,m_l)\), while \(d\) affects only the pseudospin structure. In the boundary matching below, \(M\) controls \(k_i,\kappa_i\) through the dispersion, whereas \(d\) enters via the spinor coefficients \((A_i,\bar A_i,\dots)\).

With the potential condition in this paper, as $ V_1 = V_3 = 0 $ and $ V_2 = u_0 $ for some positive value $u_0$ bigger than $E$, which is kinetic energy of incident electron, momentum and decay constant for each $i$-th area are given by
\begin{eqnarray}
   k_1 = k_3 = k = \sqrt{\frac{ 4 M E }{ 1 + 4 M \alpha }} \frac{1}{\hbar},~
   k_2 = q = \sqrt{\frac{ 4 M ( u_0 -E ) }{ 1 - 4 M \alpha }} \frac{1}{\hbar}. \nonumber
\end{eqnarray}
With translational invariance along $y$, the transverse momentum is conserved:
\[
k_y = k \sin\phi = q \sin\theta.
\]
The longitudinal components are
\[
k_{ix}=\sqrt{k_i^2-k_{iy}^2},~ \kappa_{ix} = \sqrt{k_{ix}^2+2 k_{iy}^2}.
\]
Here $\phi$ is the incidence and transmission angle in regions I and III, and $\theta$ is the refraction angle in region II.

For convenience in the following analysis, the wavefunction given in Eq.~(5) is here presented by explicitly separating its upper and lower components:
\begin{align}
&~~~~~~~~~~~~~~~~~~\Psi^{(i)}=
\begin{pmatrix}
\psi_1^{(i)}\\[2pt]
\psi_2^{(i)}
\end{pmatrix},\nonumber\\ 
\psi_1^{(i)}(x, y) = ( &a_i A_i e^{i k_{ix} x} + b_i B_i e^{-i k_{ix} x} + c_i C_i e^{\kappa_{ix} x} + d_i D_i e^{-\kappa_{ix} x} ) e^{i k_{y} y}, \nonumber
\\
\psi_2^{(i)}(x, y) = s_i ( &a_i \bar{A}_i e^{i k_{ix} x } + b_i \bar{B}_i e^{-i k_{ix} x } - c_i \bar{C}_i  e^{\kappa_{ix} x} - d_i \bar{D}_i   e^{-\kappa_{ix} x} ) e^{i k_{y} y} \nonumber
\end{align}
where
\begin{align}
&A_1 = f(\phi),~\bar{A}_1 = g(\phi),~B_1 = f(\phi),~\bar{B}_1 = g^*(\phi), C_1 = -1 + h(\phi),~\bar{C}_1 = 1 + j_1(\phi) - j_2(\phi),~ \nonumber
\\
&D_1 = \bar{D}_1=0,~A_2 = 1 - h(\theta),\bar{A}_2 = 1 - j_1(\theta) + id \sin{2\theta}, B_2 = 1 - h(\theta),~\bar{B}_2 = 1 - j_1(\theta) - id \sin{2\theta}, \nonumber
\\
&C_2 = 1 + h(\theta),~\bar{C}_2 = 1 + j_1(\theta) - j_2(\theta), D_2 = 1 + h(\theta),~\bar{D}_2 = 1 + j_1(\theta) + j_2(\theta), \nonumber
\\
&A_3 = f(\phi),~\bar{A}_3 = g(\phi),~B_3 = \bar{B}_3 = C_3 = \bar{C}_3 = 0, D_4 = -1 + h(\phi),~\bar{D}_4 = 1 + j_1(\phi) + j_2(\phi). \nonumber
\end{align}
with $\bar{d} = \left(d \sqrt{1-d^2}\right)$, $a^*$ is complex conjugate of $a$, $f(x) = \cos^{2}x+ \left( 1-2\bar{d} \right)\sin^2{x}$,
$g(x) = 1-d^2+d^2 \cos{2x} + i d \sin{2x}$,
$h(x) = 2\bar{d}\sin^2{x}$,
$j_1(x) = 2d^2 \sin^2{x}$,
$j_2(x) = 2d \sin{x}\sqrt{1 - \sin^2{x}}$ for $u_0>E$ and $s_i$ is $\operatorname{sgn}(V_i - E)$.
With this square potential barrier the spinor and its first $x$-derivative are continuous:
\[
\Psi^{(\mathrm{I})}(0,y)=\Psi^{(\mathrm{II})}(0,y),\qquad
\partial_x\Psi^{(\mathrm{I})}(0,y)=\partial_x\Psi^{(\mathrm{II})}(0,y),
\]
\[
\Psi^{(\mathrm{II})}(a,y)=\Psi^{(\mathrm{III})}(a,y),\qquad
\partial_x\Psi^{(\mathrm{II})}(a,y)=\partial_x\Psi^{(\mathrm{III})}(a,y)
\]
with potential barrier width $a$. The transmission probability $T$ can then be obtained as: 
\begin{equation}
    T = \left| \frac{a_3}{a_1} \right|^2. \nonumber
\end{equation}

\newpage
\section{The Delta Function Barrier Limit}\label{app:delta}
In this section, we consider the delta function barrier limit, where the potential is given by $V(x) = U \delta(x)$. The potential strength $U$ is defined as the product $U = u_0 a$. This definition is used to directly compare this case with the finite square potential barrier in the limit where its width $a \to 0$. A key difference from the finite barrier is that we do not consider the wavefunction within the barrier region itself. For the delta function problem, we will restrict our analysis to the isotropic case where the upper and lower bands are symmetric ($m_u = -m_l$). To facilitate an intuitive comparison with the bilayer graphene case, we will use the following Hamiltonian, which is a unitarily transformed version of the model from the main text under this condition: \begin{align}
H_0(\bm{k}) &= h_x(\mathbf{k}) \, \sigma_x
+ h_y(\mathbf{k}) \, \sigma_y
+ h_z(\mathbf{k}) \, \sigma_z, \nonumber \\
h_x(\mathbf{k}) &=  \frac{1}{4M} k_x^2 + \frac{1}{4M}(1 - 2d^2) k_y^2, \nonumber \\
h_y(\mathbf{k}) &= - \frac{1}{2M}d \sqrt{1 - d^2} \, k_y^2, \nonumber \\
h_z(\mathbf{k}) &= \frac{1}{2M} d k_x k_y. \nonumber
\end{align}
Under these conditions, we define the wavefunctions as follows. For simplicity, we will only consider the components in the $x$-direction.
\[
\Psi_1(x)=
a_1
\begin{pmatrix}
A\\
B
\end{pmatrix}
e^{i k_x x}
\;+\;
b_1
\begin{pmatrix}
A\\B^*
\end{pmatrix}
e^{-i k_x x}
\;+\;
c_1
\begin{pmatrix}
A\\[2pt]
 \,C
\end{pmatrix}
e^{\kappa_x x}.
\]
\[
\Psi_2(x)=
a_2
\begin{pmatrix}
A\\
B
\end{pmatrix}
e^{i k_x x}
\;+\;
d_2
\begin{pmatrix}
A\\[2pt]
 \,D
\end{pmatrix}
e^{-\kappa_x x}.
\]
where $\Psi_1(x)$ is the wavefunction for the region $x \le 0$, $\Psi_2(x)$ is for the region $x > 0$, $A = \cos^2\phi+\bigl(1-2d\sqrt{1-d^2}\bigr)\sin^2\phi$, $B = 1-d^2+d^2\cos(2\phi)+i d \sin(2\phi)$, $B^*$ is the complex conjugate of $B$, $C = 1 + 2 d \sin{(\phi)}(d \sin{(\phi)}-\sqrt{1+\sin{(\phi)^2}})$ and $D = 1 + 2 d \sin{(\phi)}(d \sin{(\phi)}+\sqrt{1+\sin{(\phi)^2}})$.
The system is solved by applying the boundary conditions at $x=0$.
The first boundary condition is the continuity of the wavefunction,
\[
\Psi_1(0) = \Psi_2(0)
\]
The second condition is derived by integrating the Schrödinger equation across the delta function potential. This yields a discontinuity in the derivative of the wavefunction and it is given by:
\[
\Psi_2'(0) = \Psi_1'(0) + \left( \frac{4MU}{\hbar^2} \right) \sigma_x \Psi_1(0)
\]
where $\Psi_1'(x=0)$ and $\Psi_2'(x=0)$ are the derivatives of the wavefunctions evaluated at $x=0$ from the left and right, respectively.
Solving the boundary conditions, we can get exact compact form of transmission amplitude
\begin{equation}
    t(\phi; d)=\frac{1}{\,1+iA(\phi)+d^{2} B(\phi)\,} \nonumber
\end{equation}
where
\[
A(\phi) = \rho \sec{(\phi)}, ~
B(\phi) = -\frac{2 \rho^2 \sin^2\phi (S(\phi) \cos\phi  - 2i \sin\phi)}{\cos\phi (S(\phi) - 2\rho)},~ \rho = \frac{\sqrt{M}\,u_0 a}{\sqrt{E}\,\hbar},~ S(\phi) = \sqrt{6-2 \cos{(2\phi)}}.
\]
In the limit $d = 0$, this expression reduces to the well-known transmission coefficient for free electrons in a delta potential.

\paragraph*{From $|N|^2=|D|^2$ to the closed condition in $x=\cos 2\phi$.}
Let $c\equiv\cos\phi$, $s\equiv\sin\phi$, $S\equiv S(\phi)=\sqrt{6-2\cos(2\phi)}$, and $K\equiv S-2\rho$.
With
\[
N=c\,K,\qquad
D=c\,K+i\rho\,K+2\rho^2(-cS+2is^2)s^2,
\]
we have
\[
\Re D=c\,(K-2\rho^2 S s^2),\qquad
\Im D=\rho K+4\rho^2 s^4.
\]
The perfect-transmission condition $|N|^2=|D|^2$ becomes
\begin{align}
0&=(\Re D)^2+(\Im D)^2-N^2 \nonumber\\
&=c^2\!\left[(K-2\rho^2 S s^2)^2-K^2\right]+\left(\rho K+4\rho^2 s^4\right)^2 \nonumber\\
&=c^2\!\left(-4\rho^2 S s^2 K+4\rho^4 S^2 s^4\right)+\rho^2 K^2+8\rho^3 K s^4+16\rho^4 s^8. \label{eq:Ephi} 
\end{align}

Now set $x\equiv \cos(2\phi)$ so that
\[
s^2=\frac{1-x}{2},\qquad c^2=\frac{1+x}{2},\qquad S=\sqrt{6-2x}=\sqrt{2}\,\sqrt{\,3-x\,}.
\]
Substituting these into eq.~\refeq{eq:Ephi} and dividing by $\rho^2>0$ yields
\begin{align}
0&=K^2+8\rho K\Big(\frac{1-x}{2}\Big)^{\!2}+16\rho^2\Big(\frac{1-x}{2}\Big)^{\!4}
+\frac{1+x}{2}\!\left[-4S\Big(\frac{1-x}{2}\Big)K+4\rho^2 S^2\Big(\frac{1-x}{2}\Big)^{\!2}\right] \nonumber\\
&=K^2+8\rho K\frac{(1-x)^2}{4}+16\rho^2\frac{(1-x)^4}{16}
+\frac{1+x}{2}\!\left[-2S(1-x)K+\rho^2 S^2\frac{(1-x)^2}{1}\right]. \nonumber
\end{align}
Using $K=S-2\rho$ and $S=\sqrt{2}\sqrt{3-x}$, a straightforward expansion and collection of terms in $x$ and $\rho$ gives the compact factorized form
\[
-2\left[-2\rho^2+2\sqrt{2}\,\rho\,x\sqrt{3-x}+x^3-3x^2\right]=0.
\]
Therefore, the $T=1$ condition is equivalently
\begin{equation}
\boxed{~
x\Big[x(x-3)+2\sqrt{2}\,\rho\,\sqrt{\,3-x\,}\Big]=2\rho^2,\qquad x=\cos(2\phi)\in[-1,1].\nonumber
~} \label{eq:T1final}
\end{equation}
\newpage
\section{Contour plot of the transmission probability $T_{\mathrm{max}}$}
\label{app:parameters}

\begin{figure}[h] 
    \centering
    \includegraphics[width=86mm]{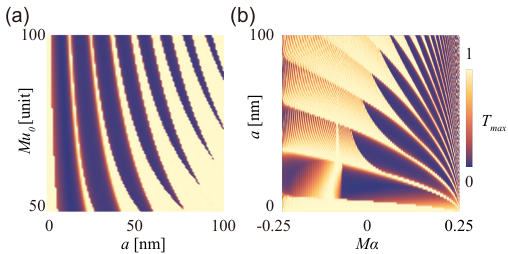}
    \caption{
    (a)[b] Contour plot of the transmission probability $T_{\mathrm{max}}$ as a function of the barrier height $M u_0$(barrier width $a$) and the barrier width $a$(mass asymmetry parameter $M \alpha$).
    }\label{fig:Tmax_contour}
\end{figure}

In Fig.~\ref{fig:Tmax_contour} we summarize how the peak value of the 
transmission varies over the parameter space. For each choice of barrier 
height $M u_0$, barrier width $a$, and mass-asymmetry parameter $M\alpha$, 
we compute the angle-resolved transmission $T(\phi)$ and define 
$T_{\mathrm{max}}$ as its maximum with respect to the incidence angle 
$\phi$ in the range $|\phi|\le\pi/2$. Panel (a) shows $T_{\mathrm{max}}$ 
in the $(M u_0,a)$ plane for fixed $M\alpha$ and $d_{\mathrm{max}}$, 
highlighting Fabry–Pérot-like resonant ridges as the barrier length 
supports constructive interference. Panel (b) displays $T_{\mathrm{max}}$ 
in the $(a,M\alpha)$ plane for fixed $M u_0$, illustrating how variations 
in the mass-asymmetry parameter reorganize the pattern of resonant and 
non-resonant regions. Taken together, these contour plots provide  
overview of the parameter dependence of $T_{\mathrm{max}}$ and complement 
the angle-resolved profiles discussed in the main text.

\newpage
\section{d part}

The parameter \( d \) in this model directly determines the Hilbert–Schmidt quantum distance between the conduction and valence band eigenstates. Specifically, the quantum distance in momentum space is defined as
\begin{equation}
d^2_{\mathrm{HS}}(\bm{k}, \bm{k}') = 1 - \left| \langle \psi_{n,\bm{k}} | \psi_{n,\bm{k}'}  \rangle \right|^2,
\end{equation}
where \( n \) denotes the band index and \( \psi_{n,\bm{k}} \) refers to the state of the \( n \)th energy band at momentum \( \bm{k} \). The maximum value of this distance across the Brillouin zone, denoted as \( d_{\mathrm{max}} \), characterizes the global quantum geometry of the system.

In our Hamiltonian, \( d \) is proportional to this maximum distance, i.e., \( d = \xi d_{\mathrm{max}} \) with \( \xi = \pm1 \). Owing to the symmetry of the system under the transformation \( d \rightarrow -d \), we focus on the range \( d \in [0, 1] \) without loss of generality. This formulation enables us to continuously interpolate between quantum geometrically trivial systems such as the free electron model (\( d = 0 \)) and maximally entangled systems like bilayer graphene (\( d = 1 \)), while keeping the energy dispersion fixed.

In particular, the maximum quantum distance between the eigenstates of the two bands is governed by \(d_{\mathrm{max}} = d\), which allows us to continuously interpolate between different quantum geometries without altering the band dispersion.

The form of this Hamiltonian ensures QBT at \(\bm{k} = 0\), with degeneracy protected by rotational symmetry. When \(d = 1\), the model reduces to the well-known low-energy Hamiltonian of bilayer graphene \((H(\bm{k})|_{d=1} = H_{\text{BLG}}(\bm{k}))\). As \(d\) is tuned from 1 to 0, the eigenstates deform while the band structure remains unchanged, allowing for an isolated study of how quantum geometry affects physical observables such as tunneling probability.

\end{document}